\documentclass[12pt,a4paper]{article}
\usepackage[text={16.5cm,24cm},centering]{geometry}
\usepackage{amsmath}
\usepackage{amsthm}
\usepackage[utf8]{inputenc}
\usepackage{tgtermes,tgheros,tgcursor}
\usepackage[varg]{newtxmath}
\usepackage{hyperref}
\hypersetup{%
 colorlinks=true,%
 linkcolor=blue,%
 citecolor=blue,
}

\numberwithin{equation}{section}

\newcommand{\pdv}[1]{\frac{\partial}{\partial #1}}

\newcommand{\del}{\partial}

\newcommand{\Zb}{\mathbb{Z}}

\newcommand{\Lcal}{\mathcal{L}}

\newcommand{\xyz}{\sharp}

\newcommand{\SCS}{S_{\mathrm{CS}}}
\newcommand{\Scp}{S_{\mathrm{couple}}}

\newcommand{\Seff}{S_{\mathrm{eff}}}
\newcommand{\gex}{h}

\begin{document}
\begin{center}
  \begin{flushright}
    OU-HET 1111
  \end{flushright}
  \vspace{5ex}
  {\LARGE \bfseries \boldmath Gapless edge modes in (4+1)-dimensional topologically massive tensor gauge theory and anomaly inflow for subsystem symmetry}\\
  \vspace{4ex}
  {\Large Satoshi Yamaguchi}\\
  \vspace{2ex}
  {\itshape Department of Physics, Graduate School of Science, 
  \\
  Osaka University, Toyonaka, Osaka 560-0043, Japan}\\
  \vspace{1ex}
  \texttt{yamaguch@het.phys.sci.osaka-u.ac.jp}\\
  \begin{abstract}
    We consider the $(4+1)$-dimensional topologically massive tensor gauge theory.  This theory is an analog of the $(2+1)$-dimensional topologically massive Maxwell-Chern-Simons theory.  If the space has a boundary, we find that a $(3+1)$-dimensional gapless theory appears at the boundary.  This gapless theory is a chiral version of the $(3+1)$-dimensional $\varphi$ theory.  This gapless theory is protected by the anomaly inflow mechanism for subsystem symmetry.  We also consider the corner of our topologically massive tensor gauge theory, and find that an infinite number of $(1+1)$-dimensional chiral bosons appear at the corner.
  \end{abstract}
\end{center}

\vspace{4ex}
\section{Introduction and summary}
Subsystem symmetries are one of the interesting generalizations of symmetry in quantum field theories \cite{Batista:2004sc,Nussinov:2006iva,Nussinov:2009zz}.
They play an important role in fracton phases \cite{Chamon:2004lew,Haah:2011drr,Nandkishore:2018sel,Pretko:2020cko}.
Such exotic systems are mostly investigated in lattice quantum mechanics \cite{Vijay:2015mka,Vijay:2016phm}. There are other approaches such as foliated quantum field theory \cite{Slagle:2020ugk,Hsin:2021mjn},  infinite component Chern-Simons theory \cite{ma2020fractonic}, and  continuous quantum field theory \cite{Pretko:2016kxt,Pretko:2016lgv,2018PhRvB..98c5111M,Bulmash:2018lid,Seiberg:2019vrp,Seiberg:2020bhn,Seiberg:2020wsg,Seiberg:2020cxy,Gorantla:2020xap}.
In this paper, we employ continuous quantum field theory approach.
More recently, subsystem symmetry and fracton phases have also been studied from the high energy physics points of views, such as supersymmetry \cite{Yamaguchi:2021qrx}, quiver gauge theory \cite{Razamat:2021jkx}, and string theory \cite{Geng:2021cmq}.

Another important phenomenon in the recent studies of the topological aspects of quantum field theories is the anomaly inflow mechanism \cite{Callan:1984sa}.
An example of this anomaly inflow is that a $(1+1)$-dimensional chiral boson appears at the boundary of the (2+1)-dimensional topologically massive Maxwell-Chern-Simons theory \cite{Deser:1981wh,Deser:1982vy}.
The anomaly of the (1+1)-dimensional chiral boson cancels to the anomaly inflow from the (2+1)-dimensional bulk, and the total system is anomaly-free.
The generalization of this example has been investigated in detail in \cite{Hsieh:2020jpj}.
The anomaly inflow mechanism lead to the modern understanding of anomalies in terms of higher dimensional symmetry protected topological (SPT) phases (see \cite{Freed:2014iua,Monnier:2019ytc} for example).

Combining these two ideas, one may find that subsystem SPT (SSPT) phases are also interesting phases of matters.
SSPT phases have been introduced and studied in many papers including \cite{You2018sspt,Devakul:2018fhz,Devakul:2019duj}.
We expect that SSPT phases and anomaly inflow are useful to understand anomalies of subsystem symmetries.

In this paper, we study an example of the SSPT phase and anomaly inflow.
We study $(4+1)$-dimensional topologically massive tensor gauge theory.
This theory is an analog of $(3+1)$-dimensional topologically massive Maxwell-Chern-Simons theory.
We explicitly construct the gapless modes localized at the boundary.
They form the chiral version of the $(3+1)$-dimensional $\varphi$ theory.
The $(3+1)$-dimensional $\varphi$-theory resembles to $(1+1)$-dimensional free scalar \cite{Gorantla:2020xap}.  Thus we expect the chiral version of the $\varphi$ theory to exist. We find it appears at the boundary of the topologically massive tensor gauge theory.

This $(4+1)$-dimensional topologically massive tensor gauge theory has subsystem symmetry called the magnetic center symmetry.
We find that the boundary chiral $\varphi$ theory is robust due to the anomaly inflow mechanism \cite{Hsieh:2020jpj} of this magnetic center symmetry.

We also consider the codimension 3 corner localized modes of the topologically massive tensor gauge theory.  We find that an infinite number of $(1+1)$-dimensional chiral bosons appear at the corner.  The anomaly inflow argument for these corner states is an interesting future problem.

Another interesting future issue is to develop a more precise formulation of the tensor Chern-Simons theory as done in \cite{Hsieh:2020jpj}.  In other words, it should be nice to have some sort of differential cohomology and differential K-theory for tensor gauge fields for subsystem symmetries.

The construction of this paper is as follows.
In Sec.~\ref{sec:codim1}, we set up the topologically massive tensor gauge theory and find the massless chiral $\varphi$ theory at the boundary of the spacetime.
In Sec.~\ref{sec:inflow}, we discuss the anomaly inflow for the magnetic center symmetry.
In Sec.~\ref{sec:corner}, we consider the corner states of the topologically massive tensor gauge theory and find an infinite number of chiral bosons.

\vspace{2ex}
Note added:  while writing this manuscript, we have noticed that the anomaly inflow mechanism of the chiral $\varphi$ theory has also been discussed in the recent paper \cite{Burnell:2021reh}.

\section{(4+1)-dimensional topologically massive tensor gauge theory on the spacetime with boundary}
\label{sec:codim1}

\subsection{Setup}

We consider here a 5-dimensional non-relativistic quantum field theory. 
The spacetime coordinates are denoted by $x^{\mu},\ \mu=0,1,2,3,5$. Here, $x^4$ is not used in the Lorentzian signature. It is reserved for Euclidean time.

It is convenient to introduce labels $A,B,C,\dots=0,\xyz,5$,
\footnote{It may become clearer why we call the $C$ field a ``tensor'' gauge field if we use label $123$ instead of $\xyz$.
However, we do not use this notation to avoid confusion when many such labels are used at the same place.}
and define the derivatives $\del_{A}$ as
\begin{align}
  \del_{0}:=\pdv{x^0},\quad \del_{5}:=\pdv{x^5}, \quad \del_{\xyz}:=\pdv{x^1}\pdv{x^2}\pdv{x^3}.
\end{align}
It is useful to notice the relation for two functions $\Phi_1(x),\ \Phi_2(x)$:
\begin{align}
 (\del_{A}\Phi_1)\Phi_2+\Phi_1 (\del_{A}\Phi_2)=(\text{total derivative}).  
\end{align}
This relation allows us to perform integration by parts in a closed spacetime.

We introduce a 5-dimensional tensor gauge field $C_{A},\ A=0, \xyz,5$.  
The gauge transformation is given by
\begin{align}
  C_{A}'=C_{A}+\del_{A}\lambda. \label{gaugetransf}
\end{align}
Here the parameter $\lambda$ has the same properties as the $\varphi$ field in the $\varphi$ theory in \cite{Seiberg:2020bhn,Gorantla:2020xap}.
For example, the periodicity $\lambda\sim \lambda+2\pi$ is imposed and certain discontinuity for $\lambda$ is allowed.

The gauge invariant field strength $G_{AB}$ is defined as
\begin{align}
  G_{AB}:=\del_{A}C_{B}-\del_{B}C_{A}.
\end{align}
We impose Lorentz symmetry in the $05$-plane, for simplicity.
The tensor Maxwell action is given by
\begin{align}
  S_{M}=\int d^5x\left[
    \frac{1}{2\gex^2}G_{05}^2+\frac{1}{2g^2}G_{\xyz 0}^2-\frac{1}{2g^2}G_{\xyz 5}^2\right],
\end{align}
where $g$ and $\gex$ are real positive coupling constants.
Since we only impose the Lorentz symmetry in the $05$-plane, there are two coupling constants.

We would also like to consider tensor Chern-Simons theory:
\begin{align}
  \SCS=\frac{k}{4\pi}\int d^5x \epsilon^{ABC}C_{A}\del_{B}C_{C},
\end{align}
where $\epsilon^{ABC}$ is a totally anti-symmetric symbol and $\epsilon^{0\xyz 5}=1$.
$k$ is an analog of the Chern-Simons level and it must be an integer for the gauge invariance as discussed in Appendix \ref{app:levelintegrality}.

We consider the theory including both the Maxwell term and the Chern-Simons term with $k=1$.
The total action is given by
\begin{align}
  S=\int d^5x\left[
    \frac{1}{2\gex^2}G_{05}^2+\frac{1}{2g^2}G_{\xyz 0}^2-\frac{1}{2g^2}G_{\xyz 5}^2+\frac{1}{4\pi}\epsilon^{ABC}C_{A}\del_{B}C_{C}
  \right].\label{MCSaction}
\end{align}
The equations of motion derived from this action are given by
\begin{align}
  &-\frac{1}{\gex^2}\del_{0}G_{05}+\frac{1}{g^2}\del_{\xyz}G_{\xyz 5}+\frac{1}{2\pi}G_{0\xyz}=0,\label{eom1}\\
 & \frac{1}{\gex^2}\del_{5}G_{05}-\frac{1}{g^2}\del_{\xyz}G_{\xyz 0}
 +\frac{1}{2\pi} G_{\xyz 5}=0,\label{eom2}\\
 &\frac{1}{g^2}\del_{0}G_{\xyz 0}-\frac{1}{g^2}\del_{5}G_{\xyz 5}-\frac{1}{2\pi}G_{05}=0.\label{eom3}
\end{align}

\subsection{Localized modes at the boundary}

Here, we consider the spacetime $x^5>0$ with boundary at $x^5=0$.
We impose the boundary condition
\begin{align}
  C_{A}|_{x_5=0}=0,
  \label{boundarycondition}
\end{align}
up to the gauge transformation at the boundary.
It is not only flat but a pure gauge.
In other words, all the Wilson loops vanish at the boundary.
Even though the theory includes Chern-Simons term and the spacetime has a boundary, the theory is gauge invariant due to the boundary condition \eqref{boundarycondition}.

We want to find the modes localized at this boundary.
To do this, we will solve the equations of motion \eqref{eom1}, \eqref{eom2}, \eqref{eom3} under the boundary condition \eqref{boundarycondition}.

We employ the ansatz
\begin{align}
  G_{\xyz 0}=0
\end{align}
even in the bulk and look for localized solutions.
Then the equations of motion \eqref{eom1}, \eqref{eom2}, \eqref{eom3} are rewritten as
\begin{align}
  &-\frac{1}{\gex^2}\del_{0}G_{05}+\frac{1}{g^2}\del_{\xyz}G_{\xyz 5}=0,\label{eom21}\\
 & \frac{1}{\gex^2}\del_{5}G_{05}
 +\frac{1}{2\pi} G_{\xyz 5}=0,\label{eom22}\\
 &-\frac{1}{g^2}\del_{5}G_{\xyz 5}-\frac{1}{2\pi}G_{05}=0,\label{eom23}  
\end{align} 
respectively.  We solve \eqref{eom23} for $G_{05}$ and substitute it into \eqref{eom22}.  Then we obtain
\begin{align}
  \del_{5}^2 G_{\xyz 5}=m^2 G_{\xyz 5},\qquad m:=\frac{\gex g}{2\pi}>0.
\end{align}
For a normalizable solution, this equation implies
\begin{align}
  \del_{5}G_{\xyz 5}=-mG_{\xyz 5}. \label{tempsolveeom1}
\end{align}
In other words, the $x^5$ dependence of $G_{\xyz 5}$ is $G_{\xyz 5}\propto e^{-mx^5}$.  Similarly we solve \eqref{eom22} for $G_{\xyz 5}$ and substitute it into \eqref{eom23}.  Then, taking the normalizability into account, we obtain
\begin{align}
  \del_{5}G_{0 5}=-mG_{0 5}.\label{tempsolveeom2}
\end{align}
By substituting \eqref{tempsolveeom1} and \eqref{tempsolveeom2} into \eqref{eom22} and \eqref{eom23}, we obtain
\begin{align}
  G_{05}-\alpha G_{\xyz 5}=0,\quad \alpha:=\frac{\gex}{g}.
  \label{constraint}
\end{align}

We choose the $C_5=0$ gauge. Since $C_{0},C_{\xyz}$ are a pure gauge at the boundary $x_5=0$, they are written using a $\varphi$ field with the periodicity $\varphi\sim \varphi+2\pi$ of the $\varphi$ theory at the boundary as
\begin{align}
  C_{0}|_{x^5=0}=\del_{0}\varphi,\quad C_{\xyz}|_{x^5=0}=\del_{\xyz}\varphi.
\end{align}
The $x^5$ dependence is determined by Eqs.~\eqref{tempsolveeom1}, \eqref{tempsolveeom2}, and we obtain
\begin{align}
  C_{0}=\del_{0}\varphi e^{-mx^5},\quad C_{\xyz}=\del_{\xyz}\varphi e^{-mx^5}.
\end{align}

Eq.~\eqref{eom21} reads
\begin{align}
  \del_{0}^2\varphi-\alpha^2 \del_{\xyz}^2\varphi=0.
\end{align}
This is the equation of motion of the $\varphi$ theory \cite{Gorantla:2020xap} (see also \cite{Yamaguchi:2021qrx}). Actually, our $\varphi$ field must satisfy a stronger constraint from Eq.~\eqref{constraint}:
\begin{align}
  \del_{0}\varphi-\alpha \del_{\xyz}\varphi=0. \label{boundarylocalizedchiralmodes}
\end{align}
This constraint implies that these boundary localized modes form the chiral $\varphi$ theory.

Here let us comment on the Chern-Simons level.  In this paper, we choose $k=+1$ and obtain the boundary localized chiral modes.  If we consider the case with $k=-1$, we obtain boundary localized anti-chiral modes:
$\del_{0}\varphi+\alpha \del_{\xyz}\varphi=0.$
If we consider $|k|>1$, the story is quite different.  The bulk theory is not in an SSPT phase anymore but has some topological degrees of freedom.  Moreover, we cannot impose the Dirichlet boundary condition \eqref{boundarycondition}.  It is an interesting issue to study this case in more detail.

\section{Anomaly inflow}
\label{sec:inflow}
In the previous section, we have shown that the $(3+1)$-dimensional chiral $\varphi$ theory appears at the boundary of the $(4+1)$-dimensional topologically massive tensor gauge theory.
In this section, we want to show that this boundary theory is robust due to the anomaly inflow mechanism for subsystem symmetry.

In this subsection, we use the Euclidean formulation.
The coordinates are $x^{\mu},\ \mu=1,\dots,5$.
$x^4$ is Euclidean time.
It is convenient to use labels $A,B,C=\xyz,4,5$.
The symbol $\epsilon^{ABC}$ is totally anti-symmetric and $\epsilon^{\xyz 4 5}=1$.
The Euclidean action of the topologically massive tensor gauge theory is given by
\begin{align}
  S_{E}=\int d^5x\left[
    \frac{1}{2\gex^2}G_{45}^2+\frac{1}{2g^2}G_{\xyz 4}^2+\frac{1}{2g^2}G_{\xyz 5}^2+\frac{i}{4\pi}\epsilon^{ABC}C_{A}\del_{B}C_{C}
  \right].\label{MCSEuclideanaction}
\end{align}

\subsection{Magnetic center symmetry}
We have a global symmetry called ``magnetic center symmetry'' in the topologically massive tensor gauge theory \eqref{MCSaction}.  The current of this symmetry is given by
\begin{align}
  M^{A}:= \frac{i}{2\pi}\epsilon^{ABC}\del_{B}C_{C}.
\end{align}
This current satisfies the conservation law:
\begin{align}
  \del_{A}M^{A}=0.
\end{align}
This magnetic center symmetry is a subsystem symmetry.  This magnetic center symmetry is preserved even if we introduce boundary at $x^5=0$ with the boundary condition \eqref{boundarycondition}.

Let us consider gauging this symmetry by introducing background gauge field $A_{A}$ and adding the term
\begin{align}
  \Scp=\int d^5 x  A_{A}M^{A}=\frac{i}{2\pi}\int d^5 x \epsilon^{ABC}A_{A}\del_{B}C_{C}. \label{coupling term}
\end{align}
The gauge transformation is given by
\begin{align}
  A'_{A}=A_{A}+\del_{A}\rho, \label{infinitesimalgaugetransf}
\end{align}
where $\rho$ is the parameter of the gauge transformation. In this paper, we only consider infinitesimal gauge transformation for the background gauge field, and thus $\rho$ is an infinitesimal function on the spacetime.  The term \eqref{coupling term} is invariant under the infinitesimal gauge transformation \eqref{infinitesimalgaugetransf} if the spacetime does not have a boundary.

\subsection{Effective theory of the background gauge field}
Let us consider the topologically massive tensor gauge theory coupled to the background magnetic center symmetry gauge field.  The partition function with the background gauge field is given by
\begin{align}
  Z[A]=\int DCe^{-S_{E}(C)-\Scp(C,A)}=:|Z[A]|e^{-\Seff(A)}.
\end{align}
The phase functional $e^{-\Seff}$ characterizes the SSPT phase.
We only consider the low energy limit and therefore ignore the Maxwell term.
We can rewrite the action as
\begin{align}
  S_{E}+\Scp=&
  \frac{i}{4\pi}\int d^5x \epsilon^{ABC}(C_{A}+A_{A})\del_{B}(C_{C}+A_{C})
  -\frac{i}{4 \pi}\int d^5x \epsilon^{ABC} A_{A}\del_{B}A_{C}\nonumber\\
  &+\frac{i}{4\pi}\int d^5x \epsilon^{ABC}\del_{B}(
    A_{A}C_{C}).
\end{align}
The first term will disappear if we redefine the $C$ field and integrate it out.  The second term remains and becomes a part of $\Seff$.
The third term is a total derivative and therefore we can ignore it at least perturbatively.  This is enough since we are considering perturbative anomaly.
Then, $\Seff$ is given by
\begin{align}
  \Seff(A)=-\frac{i}{4 \pi}\int d^5x \epsilon^{ABC} A_{A}\del_{B}A_{C}.
 \label{Seff}
\end{align}
This is gauge invariant if the spacetime is closed.

Once we consider the spacetime with boundary, the effective action \eqref{Seff} is not gauge invariant anymore.  Let us consider the spacetime $x^5>0$ with boundary $x^5=0$.  Then the gauge transformation of \eqref{Seff} is given by
\begin{align}
  \delta\Seff(A)=\frac{i}{4 \pi}\int_{x^5=0} d^4x \epsilon^{5BC} \rho\del_{B}A_{C}.
\end{align}
This should cancel the anomaly of the chiral $\varphi$ theory.  Therefore, the chiral $\varphi$ theory at the boundary is robust due to this anomaly inflow mechanism.

\section{Localized solution at the corner}
\label{sec:corner}

In this section, we consider the topologically massive tensor gauge theory \eqref{MCSaction} in the spacetime $x^i>0,\ i=1,2,3$.  Here, we go back to the Lorentzian signature.

\subsection{Boundary conditions}
First, we discuss the boundary conditions.  We impose the boundary condition such that the gauge symmetry \eqref{gaugetransf} is preserved.  The gauge transformation of the action \eqref{MCSaction} is calculated as
\begin{align}
  S'-S&=\frac{1}{4\pi}\int d^5 x \Delta \Lcal,\\
  \Delta \Lcal&=\epsilon^{ABC}\left[(C_{A}+\del_{A}\lambda)\del_{B}(C_C+\del_{C}\lambda)
  -C_{A}\del_{B}C_{C}
  \right]\nonumber\\
  &=\epsilon^{ABC}\del_{A}\lambda\del_{B}C_C
  =\del_{0}\lambda G_{\xyz 5}+\del_{5}\lambda G_{0\xyz}+\del_{\xyz}\lambda G_{50}\nonumber\\
  &=\del_{0}(\lambda G_{\xyz 5})+\del_{5}(\lambda G_{0\xyz})
  +\del_{\xyz}\lambda G_{50}+\lambda  \del_{\xyz} G_{50}.
\end{align}
In the last expression, the first and second total derivative terms vanish since our spacetime does not have a boundary in $x^0,x^5$ directions.  
Let us look at the third and fourth terms more carefully:
\begin{align}
  \del_{\xyz}\lambda G_{50}+\lambda  \del_{\xyz} G_{50}
  =\del_{1}(\del_2\del_3\lambda G_{50})
  -\del_{2}(\del_3\lambda \del_1 G_{50})
  +\del_{3}(\lambda \del_1\del_2 G_{50}).
\end{align}
Thus the boundary term at $x^1=0$ becomes
\begin{align}
  S'-S
  &=-\frac{1}{4\pi}\int_{x^1=0} d x^0 dx^2 dx^3 dx^5 (\del_2\del_3\lambda G_{50})+\cdots\nonumber \\
  &=-\frac{1}{4\pi}\int_{x^1=0} d x^0 dx^2 dx^3 dx^5 [\lambda \del_2 \del_3 G_{50}  -\del_3(\lambda \del_2 G_{50})+\del_2(\del_3 \lambda G_{50})]+\cdots.\label{tempgaugeboundary}
\end{align}
Here $\cdots$ represents the other boundary terms at $x^2=0$, $x^3=0$ and corners.  The first term in the integrand must vanish to be gauge invariant.  Therefore we impose the boundary condition
\begin{align}
  \del_2\del_3 G_{50}=0\quad \text{at } x^1=0.
\end{align}
The third term of the integrand in Eq.~\eqref{tempgaugeboundary} contributes to the corner $x^1=x^2=0$.  In order to eliminate this contribution, we impose the boundary condition
\begin{align}
  \del_3 G_{50}=0\quad \text{at } x^1=x^2=0.
\end{align}
Here we perform integration by parts again.  Finally, we consider the corner $x^1=x^2=x^3=0$.  The boundary term at this corner is
\begin{align}
  S'-S
  =-\frac{1}{4\pi}\int_{x^1=x^2=x^3=0} d x^0 dx^5 \lambda G_{50}+\cdots.
\end{align}
Since $\lambda$ may have winding, we impose not only $G_{50}=0$ but $(C_0,C_5)$ is a pure gauge in order to be gauge invariant.

Let us summarize these boundary conditions including permutations of $x^1,x^2,x^3$.  Let $i,j,k=1,2,3$ all different labels.  Then we impose the boundary conditions:
\begin{align}
  &\del_j\del_k G_{50}=0\quad \text{at } x^i=0,\\
  &\del_k G_{50}=0\quad \text{at } x^i=x^j=0,\\
  &(C_0,C_5) \text{ is a pure gauge at } x^1=x^2=x^3=0.
\end{align}
The gauge symmetry is preserved due to these boundary conditions.

\subsection{Corner localized modes}
In this setup, we want to find the modes localized at the corner $x^1=x^2=x^3=0$.
We employ an ansatz $G_{05}=0$ in the whole spacetime.
Then the equations of motion \eqref{eom1}, \eqref{eom2}, \eqref{eom3} become
\begin{align}
&  \frac{1}{g^2} \del_{\xyz}G_{\xyz 5}-\frac{1}{2\pi} G_{\xyz 0}=0,\label{eom31}\\
&  -\frac{1}{g^2} \del_{\xyz}G_{\xyz 0}+\frac{1}{2\pi}G_{\xyz 5}=0,\label{eom32}\\
& \del_{0}G_{\xyz 0}-\del_{5} G_{\xyz 5}=0.\label{eom33}
\end{align}
By solving \eqref{eom32} in terms of $G_{\xyz 5}$ and substituting it into \eqref{eom31}, we find
\begin{align}
  \del_{\xyz}^2 G_{\xyz 0}=\frac{g^4}{(2\pi)^2} G_{\xyz 0}.
\end{align}
A normalizable solution must satisfy
\begin{align}
  \del_{\xyz}G_{\xyz 0}=-\frac{g^2}{2\pi}G_{\xyz 0}.
\end{align}
By solving \eqref{eom31} in terms of $G_{\xyz 0}$ and substituting it into \eqref{eom32}, we find that a normalizable solution satisfies
\begin{align}
  \del_{\xyz}G_{\xyz 5}=-\frac{g^2}{2\pi}G_{\xyz 5}.  
\end{align}
We substitute these equations into \eqref{eom31} and \eqref{eom32}, and obtain
\begin{align}
  G_{\xyz 5}+G_{\xyz 0}=0. \label{tempsolveeom3}
\end{align}
We choose the $C_{\xyz}=0$ gauge and find the solutions:
\begin{align}
  &C_{0}=\del_{0}\phi(x^0,x^5)\exp(-m_1x^1-m_2x^2-m_3x^3),\quad
  C_{5}=\del_{5}\phi(x^0,x^5)\exp(-m_1x^1-m_2x^2-m_3x^3),
\end{align}
where $m_1,m_2,m_3$ are real positive constants which satisfy
\begin{align}
  m_1,m_2,m_3>0,\quad m_1m_2m_3=\frac{g^2}{2\pi}.  \label{m}
\end{align}
Eq.~\eqref{tempsolveeom3} leads to a constraint for $\phi(x^0,x^5)$:
\begin{align}
\del_{0}\phi+\del_5\phi=0.  
\end{align}
This constraint implies that $\phi$ is a $(1+1)$-dimensional chiral boson.

Since there are an infinite number of possible $m_1,m_2,m_3$ satisfying \eqref{m}, we conclude that there are an infinite number of chiral bosons at the corner.

\subsection*{Acknowledgement}
The author would like to thank Sinya Aoki, Hidenori Fukaya, Yosuke Imamura, Katsushi Ito, Naoki Kawai, Yoshiyuki Matsuki, Shin'ichi Nojiri, Takuya Okuda, Kantaro Omori, Tetsuya Onogi, Tadakatsu Sakai, Sotaro Sugishita, Juven Wang, Shing-Tung Yau and Yutaka Yoshida for useful discussions and comments.
The author would also thank the Yukawa Institute for Theoretical Physics at Kyoto University.
Discussions during the YITP workshop YITP-W-21-04 on ``Strings and Fields 2021'' were useful to complete this work. 
This work was supported in part by JSPS KAKENHI Grant Number 21K03574.

\appendix

\section{Level of the tensor Chern-Simons theory}
\label{app:levelintegrality}

In this appendix, we argue that $k$ must be an integer in the tensor Chern-Simons theory:
\begin{align}
  \SCS=\frac{ik}{4\pi}\int d^5x \epsilon^{ABC}C_{A}\del_{B}C_{C}.
\end{align}
We follow the analogous argument to \cite{Witten:2015aoa}.
We only consider the spacetime manifold $T^5$ in which all $x^M,\ M=1,2,3,4,5$ are periodic.  We consider gauge invariance of the theory in the following two gauge field configurations.
\begin{enumerate}
  \item One unit of tensor electric flux in $1234$-space.
  \item One unit of usual electric flux in $45$-plane.
\end{enumerate}

Let us start with the first configuration.  This implies
\begin{align}
  \int d^4 x(\del_{\xyz}C_{4}-\del_{4}C_{\xyz}) = 2\pi.
\end{align}
For example, we can choose $C_{\xyz},C_4$ as
\begin{align}
  &C_{4}=0,\qquad C_{\xyz}=-\del_{\xyz}f(x^1,x^2,x^3) \frac{x^4}{\ell_4}, \\
  &f(x^1,x^2,x^3):=2\pi\Bigg[
    \frac{x^1x^2x^3}{\ell_1 \ell_2 \ell_3}
    -\frac{x^1x^2}{\ell_1\ell_2}\theta(x^3)
    -\frac{x^2x^3}{\ell_2\ell_3}\theta(x^1)
    -\frac{x^3x^1}{\ell_3\ell_1}\theta(x^2)\nonumber\\
    &\qquad\qquad\qquad\qquad
    +\frac{x^1}{\ell_1}\theta(x^2)\theta(x^3)
    +\frac{x^2}{\ell_2}\theta(x^3)\theta(x^1)
    +\frac{x^3}{\ell_3}\theta(x^1)\theta(x^2)
  \Bigg],
\end{align}
where $\ell_i$ denotes the periodicity of $x^i$ as $x^i\sim x^i+\ell_i$, and $\theta(x)$ denotes the Heaviside step function. This $f$ is a winding number $1$ field configuration of the $\varphi$ theory \cite{Gorantla:2020xap}.
We also introduce flat $C_{5}$ with a parameter $u$ given by
\begin{align}
  \int dx_5 C_5 = 2\pi u.
\end{align}
We continuously change $u$ from $0$ to $1$ and see how $S_{CS}$ changes by this deformation.
Notice that configurations $u=0,1$ are connected by a gauge transformation and they must be identical up to $2\pi i \Zb$ in order to have gauge invariance.
It is also useful to notice that by an arbitrary small change of the gauge field $\delta C_{A}$, the change of the action is written as
\begin{align}
  \delta \SCS=\frac{ik}{2\pi}\int d^5x \epsilon^{ABC}\delta C_{A}\del_{B}C_{C}.
\end{align}
By using this equation, we obtain
\begin{align}
  \del_{u}\SCS
  &=\frac{ik}{2\pi}\int d^5x \epsilon^{ABC}\del_{u} C_{A}\del_{B}C_{C}
  =\frac{ik}{2\pi}\int d^5x \epsilon^{5ab}\del_{u} C_{5}\del_{a}C_{b}\nonumber\\
  &=\frac{ik}{2\pi}\int d^4x \epsilon^{ab}\del_{a}C_{b}\; \del_{u}\int dx^5 C_5
  =2\pi i k,\qquad (a,b=\xyz,4),
\end{align}
and therefore
\begin{align}
  \SCS|_{u=1}-\SCS|_{u=0}=\int_{0}^{1}du \del_{u}\SCS=2\pi i k.
\end{align}
As a result, we find that $k$ must be an integer.

Next, let us consider the second configuration. In this configuration,
\begin{align}
  \int dx^4dx^5(\del_4 C_5 - \del_5 C_4) = 2\pi  
\end{align}
is satisfied.  For example, we can choose
\begin{align}
  C_{4}=0,\quad C_{5}=\frac{2\pi}{\ell_{4}\ell_{5}}x^4.
\end{align}
Let us consider a gauge transformation parameter $\lambda_1$ which only depends on $x_1,x_2,x_3$ and satisfies
\begin{align}
  \int dx^1dx^2dx^3 \del_{\xyz}\lambda_1=2\pi.
\end{align}
This is a $\varphi$ field configuration with winding number 1 \cite{Gorantla:2020xap}.
We consider $C_{\xyz}$ with a real parameter $v$:
\begin{align}
  C_{\xyz}= v \del_{\xyz} \lambda_1.
\end{align}
The configurations $v=0,1$ are connected by a gauge transformation,
and thus they must be identical up to $2\pi i \Zb$.  By the same argument as the first configuration, we find that $k$ must be an integer.

\bibliographystyle{utphys}
\bibliography{ref}

\providecommand{\href}[2]{#2}\begingroup\raggedright\begin{thebibliography}{10}

\bibitem{Batista:2004sc}
C.~D. Batista and Z.~Nussinov, ``{Generalized Elitzur's theorem and dimensional
  reduction},'' \href{http://dx.doi.org/10.1103/PhysRevB.72.045137}{{\em Phys.
  Rev. B} {\bfseries 72} (2005) 045137},
  \href{http://arxiv.org/abs/cond-mat/0410599}{{\ttfamily
  arXiv:cond-mat/0410599}}.

\bibitem{Nussinov:2006iva}
Z.~Nussinov and G.~Ortiz, ``{Sufficient symmetry conditions for Topological
  Quantum Order},'' \href{http://dx.doi.org/10.1073/pnas.0803726105}{{\em Proc.
  Nat. Acad. Sci.} {\bfseries 106} (2009) 16944--16949},
  \href{http://arxiv.org/abs/cond-mat/0605316}{{\ttfamily
  arXiv:cond-mat/0605316}}.

\bibitem{Nussinov:2009zz}
Z.~Nussinov and G.~Ortiz, ``{A symmetry principle for topological quantum
  order},'' \href{http://dx.doi.org/10.1016/j.aop.2008.11.002}{{\em Annals
  Phys.} {\bfseries 324} (2009) 977--1057},
  \href{http://arxiv.org/abs/cond-mat/0702377}{{\ttfamily
  arXiv:cond-mat/0702377}}.

\bibitem{Chamon:2004lew}
C.~Chamon, ``{Quantum Glassiness},''
  \href{http://dx.doi.org/10.1103/physrevlett.94.040402}{{\em Phys. Rev. Lett.}
  {\bfseries 94} no.~4, (2005) 040402},
  \href{http://arxiv.org/abs/cond-mat/0404182}{{\ttfamily
  arXiv:cond-mat/0404182}}.

\bibitem{Haah:2011drr}
J.~Haah, ``{Local stabilizer codes in three dimensions without string logical
  operators},'' \href{http://dx.doi.org/10.1103/physreva.83.042330}{{\em Phys.
  Rev. A} {\bfseries 83} no.~4, (2011) 042330},
  \href{http://arxiv.org/abs/1101.1962}{{\ttfamily arXiv:1101.1962
  [quant-ph]}}.

\bibitem{Nandkishore:2018sel}
R.~M. Nandkishore and M.~Hermele, ``{Fractons},''
  \href{http://dx.doi.org/10.1146/annurev-conmatphys-031218-013604}{{\em Ann.
  Rev. Condensed Matter Phys.} {\bfseries 10} (2019) 295--313},
  \href{http://arxiv.org/abs/1803.11196}{{\ttfamily arXiv:1803.11196
  [cond-mat.str-el]}}.

\bibitem{Pretko:2020cko}
M.~Pretko, X.~Chen, and Y.~You, ``{Fracton Phases of Matter},''
  \href{http://dx.doi.org/10.1142/S0217751X20300033}{{\em Int. J. Mod. Phys. A}
  {\bfseries 35} no.~06, (2020) 2030003},
  \href{http://arxiv.org/abs/2001.01722}{{\ttfamily arXiv:2001.01722
  [cond-mat.str-el]}}.

\bibitem{Vijay:2015mka}
S.~Vijay, J.~Haah, and L.~Fu, ``{A New Kind of Topological Quantum Order: A
  Dimensional Hierarchy of Quasiparticles Built from Stationary Excitations},''
  \href{http://dx.doi.org/10.1103/PhysRevB.92.235136}{{\em Phys. Rev. B}
  {\bfseries 92} no.~23, (2015) 235136},
  \href{http://arxiv.org/abs/1505.02576}{{\ttfamily arXiv:1505.02576
  [cond-mat.str-el]}}.

\bibitem{Vijay:2016phm}
S.~Vijay, J.~Haah, and L.~Fu, ``{Fracton Topological Order, Generalized Lattice
  Gauge Theory and Duality},''
  \href{http://dx.doi.org/10.1103/PhysRevB.94.235157}{{\em Phys. Rev. B}
  {\bfseries 94} no.~23, (2016) 235157},
  \href{http://arxiv.org/abs/1603.04442}{{\ttfamily arXiv:1603.04442
  [cond-mat.str-el]}}.

\bibitem{Slagle:2020ugk}
K.~Slagle, ``{Foliated Quantum Field Theory of Fracton Order},''
  \href{http://dx.doi.org/10.1103/PhysRevLett.126.101603}{{\em Phys. Rev.
  Lett.} {\bfseries 126} no.~10, (2021) 101603},
  \href{http://arxiv.org/abs/2008.03852}{{\ttfamily arXiv:2008.03852
  [hep-th]}}.

\bibitem{Hsin:2021mjn}
P.-S. Hsin and K.~Slagle, ``{Comments on foliated gauge theories and dualities
  in 3+1d},'' \href{http://dx.doi.org/10.21468/SciPostPhys.11.2.032}{{\em
  SciPost Phys.} {\bfseries 11} no.~2, (2021) 032},
  \href{http://arxiv.org/abs/2105.09363}{{\ttfamily arXiv:2105.09363
  [cond-mat.str-el]}}.

\bibitem{ma2020fractonic}
X.~Ma, W.~Shirley, M.~Cheng, M.~Levin, J.~McGreevy, and X.~Chen, ``Fractonic
  order in infinite-component chern-simons gauge theories,''
  \href{http://arxiv.org/abs/2010.08917}{{\ttfamily arXiv:2010.08917
  [cond-mat.str-el]}}.

\bibitem{Pretko:2016kxt}
M.~Pretko, ``{Subdimensional Particle Structure of Higher Rank U(1) Spin
  Liquids},'' \href{http://dx.doi.org/10.1103/PhysRevB.95.115139}{{\em Phys.
  Rev. B} {\bfseries 95} no.~11, (2017) 115139},
  \href{http://arxiv.org/abs/1604.05329}{{\ttfamily arXiv:1604.05329
  [cond-mat.str-el]}}.

\bibitem{Pretko:2016lgv}
M.~Pretko, ``{Generalized Electromagnetism of Subdimensional Particles: A Spin
  Liquid Story},'' \href{http://dx.doi.org/10.1103/PhysRevB.96.035119}{{\em
  Phys. Rev. B} {\bfseries 96} no.~3, (2017) 035119},
  \href{http://arxiv.org/abs/1606.08857}{{\ttfamily arXiv:1606.08857
  [cond-mat.str-el]}}.

\bibitem{2018PhRvB..98c5111M}
H.~{Ma}, M.~{Hermele}, and X.~{Chen}, ``{Fracton topological order from the
  Higgs and partial-confinement mechanisms of rank-two gauge theory},''
  \href{http://dx.doi.org/10.1103/PhysRevB.98.035111}{{\em Phys. Rev. B}
  {\bfseries 98} no.~3, (July, 2018) 035111},
  \href{http://arxiv.org/abs/1802.10108}{{\ttfamily arXiv:1802.10108
  [cond-mat.str-el]}}.

\bibitem{Bulmash:2018lid}
D.~Bulmash and M.~Barkeshli, ``{The Higgs Mechanism in Higher-Rank Symmetric
  $U(1)$ Gauge Theories},''
  \href{http://dx.doi.org/10.1103/PhysRevB.97.235112}{{\em Phys. Rev. B}
  {\bfseries 97} no.~23, (2018) 235112},
  \href{http://arxiv.org/abs/1802.10099}{{\ttfamily arXiv:1802.10099
  [cond-mat.str-el]}}.

\bibitem{Seiberg:2019vrp}
N.~Seiberg, ``{Field Theories With a Vector Global Symmetry},''
  \href{http://dx.doi.org/10.21468/SciPostPhys.8.4.050}{{\em SciPost Phys.}
  {\bfseries 8} no.~4, (2020) 050},
  \href{http://arxiv.org/abs/1909.10544}{{\ttfamily arXiv:1909.10544
  [cond-mat.str-el]}}.

\bibitem{Seiberg:2020bhn}
N.~Seiberg and S.-H. Shao, ``{Exotic Symmetries, Duality, and Fractons in
  2+1-Dimensional Quantum Field Theory},''
  \href{http://arxiv.org/abs/2003.10466}{{\ttfamily arXiv:2003.10466
  [cond-mat.str-el]}}.

\bibitem{Seiberg:2020wsg}
N.~Seiberg and S.-H. Shao, ``{Exotic $U(1)$ Symmetries, Duality, and Fractons
  in 3+1-Dimensional Quantum Field Theory},''
  \href{http://dx.doi.org/10.21468/SciPostPhys.9.4.046}{{\em SciPost Phys.}
  {\bfseries 9} no.~4, (2020) 046},
  \href{http://arxiv.org/abs/2004.00015}{{\ttfamily arXiv:2004.00015
  [cond-mat.str-el]}}.

\bibitem{Seiberg:2020cxy}
N.~Seiberg and S.-H. Shao, ``{Exotic $\mathbb{Z}_N$ Symmetries, Duality, and
  Fractons in 3+1-Dimensional Quantum Field Theory},''
  \href{http://dx.doi.org/10.21468/SciPostPhys.10.1.003}{{\em SciPost Phys.}
  {\bfseries 10} (2021) 003}, \href{http://arxiv.org/abs/2004.06115}{{\ttfamily
  arXiv:2004.06115 [cond-mat.str-el]}}.

\bibitem{Gorantla:2020xap}
P.~Gorantla, H.~T. Lam, N.~Seiberg, and S.-H. Shao, ``{More Exotic Field
  Theories in 3+1 Dimensions},''
  \href{http://dx.doi.org/10.21468/SciPostPhys.9.5.073}{{\em SciPost Phys.}
  {\bfseries 9} (2020) 073}, \href{http://arxiv.org/abs/2007.04904}{{\ttfamily
  arXiv:2007.04904 [cond-mat.str-el]}}.

\bibitem{Yamaguchi:2021qrx}
S.~Yamaguchi, ``{Supersymmetric quantum field theory with exotic symmetry in
  3+1 dimensions and fermionic fracton phases},''
  \href{http://arxiv.org/abs/2102.04768}{{\ttfamily arXiv:2102.04768
  [hep-th]}}.

\bibitem{Razamat:2021jkx}
S.~S. Razamat, ``{Quivers and Fractons},''
  \href{http://dx.doi.org/10.1103/PhysRevLett.127.141603}{{\em Phys. Rev.
  Lett.} {\bfseries 127} no.~14, (2021) 141603},
  \href{http://arxiv.org/abs/2107.06465}{{\ttfamily arXiv:2107.06465
  [hep-th]}}.

\bibitem{Geng:2021cmq}
H.~Geng, S.~Kachru, A.~Karch, R.~Nally, and B.~C. Rayhaun, ``{Fractons and
  exotic symmetries from branes},''
  \href{http://dx.doi.org/10.1002/prop.202100133}{{\em Fortsch. Phys.}
  {\bfseries 2021} (8, 2021) 2100133},
  \href{http://arxiv.org/abs/2108.08322}{{\ttfamily arXiv:2108.08322
  [hep-th]}}.

\bibitem{Callan:1984sa}
C.~G. Callan, Jr. and J.~A. Harvey, ``{Anomalies and Fermion Zero Modes on
  Strings and Domain Walls},''
  \href{http://dx.doi.org/10.1016/0550-3213(85)90489-4}{{\em Nucl. Phys. B}
  {\bfseries 250} (1985) 427--436}.

\bibitem{Deser:1981wh}
S.~Deser, R.~Jackiw, and S.~Templeton, ``{Topologically Massive Gauge
  Theories},'' \href{http://dx.doi.org/10.1016/0003-4916(82)90164-6}{{\em
  Annals Phys.} {\bfseries 140} (1982) 372--411}. [Erratum: Annals Phys. 185,
  406 (1988)].

\bibitem{Deser:1982vy}
S.~Deser, R.~Jackiw, and S.~Templeton, ``{Three-Dimensional Massive Gauge
  Theories},'' \href{http://dx.doi.org/10.1103/PhysRevLett.48.975}{{\em Phys.
  Rev. Lett.} {\bfseries 48} (1982) 975--978}.

\bibitem{Hsieh:2020jpj}
C.-T. Hsieh, Y.~Tachikawa, and K.~Yonekura, ``{Anomaly inflow and $p$-form
  gauge theories},'' \href{http://arxiv.org/abs/2003.11550}{{\ttfamily
  arXiv:2003.11550 [hep-th]}}.

\bibitem{Freed:2014iua}
D.~S. Freed, ``{Anomalies and Invertible Field Theories},''
  \href{http://dx.doi.org/10.1090/pspum/088/01462}{{\em Proc. Symp. Pure Math.}
  {\bfseries 88} (2014) 25--46},
  \href{http://arxiv.org/abs/1404.7224}{{\ttfamily arXiv:1404.7224 [hep-th]}}.

\bibitem{Monnier:2019ytc}
S.~Monnier, ``{A Modern Point of View on Anomalies},''
  \href{http://dx.doi.org/10.1002/prop.201910012}{{\em Fortsch. Phys.}
  {\bfseries 67} no.~8-9, (2019) 1910012},
  \href{http://arxiv.org/abs/1903.02828}{{\ttfamily arXiv:1903.02828
  [hep-th]}}.

\bibitem{You2018sspt}
Y.~You, T.~Devakul, F.~J. Burnell, and S.~L. Sondhi, ``Subsystem symmetry
  protected topological order,''
  \href{http://dx.doi.org/10.1103/physrevb.98.035112}{{\em Phys. Rev. B}
  {\bfseries 98} no.~3, (Jul, 2018) },
  \href{http://arxiv.org/abs/1803.02369}{{\ttfamily arXiv:1803.02369}}.
  \url{http://dx.doi.org/10.1103/PhysRevB.98.035112}.

\bibitem{Devakul:2018fhz}
T.~Devakul, D.~J. Williamson, and Y.~You, ``{Classification of subsystem
  symmetry-protected topological phases},''
  \href{http://dx.doi.org/10.1103/PhysRevB.98.235121}{{\em Phys. Rev. B}
  {\bfseries 98} no.~23, (2018) 235121},
  \href{http://arxiv.org/abs/1808.05300}{{\ttfamily arXiv:1808.05300
  [cond-mat.str-el]}}.

\bibitem{Devakul:2019duj}
T.~Devakul, W.~Shirley, and J.~Wang, ``{Strong planar subsystem
  symmetry-protected topological phases and their dual fracton orders},''
  \href{http://dx.doi.org/10.1103/PhysRevResearch.2.012059}{{\em Phys. Rev.
  Res.} {\bfseries 2} no.~1, (2020) 012059},
  \href{http://arxiv.org/abs/1910.01630}{{\ttfamily arXiv:1910.01630
  [cond-mat.str-el]}}.

\bibitem{Burnell:2021reh}
F.~J. Burnell, T.~Devakul, P.~Gorantla, H.~T. Lam, and S.-H. Shao, ``{Anomaly
  Inflow for Subsystem Symmetries},''
  \href{http://arxiv.org/abs/2110.09529}{{\ttfamily arXiv:2110.09529
  [cond-mat.str-el]}}.

\bibitem{Witten:2015aoa}
E.~Witten, ``{Three lectures on topological phases of matter},''
  \href{http://dx.doi.org/10.1393/ncr/i2016-10125-3}{{\em Riv. Nuovo Cim.}
  {\bfseries 39} no.~7, (2016) 313--370},
  \href{http://arxiv.org/abs/1510.07698}{{\ttfamily arXiv:1510.07698
  [cond-mat.mes-hall]}}.

\end{thebibliography}\endgroup
\end{document}